\documentclass[letterpaper]{article}

\usepackage{parskip}
\usepackage{graphicx}
\usepackage{amsmath}
\usepackage{caption}
\usepackage{chngpage}
\usepackage{amssymb}
\usepackage{textgreek}

\begin{document}
\title{Do City Borders Constrain Ethnic Diversity?
}
\author{Scott W. Hegerty\\
Department of Economics\\
Northeastern Illinois University\\
Chicago, IL 60625\\S-Hegerty@neiu.edu
}
\date{\ July 28, 2020}
\maketitle

\begin{abstract}

U.S. metropolitan areas, particularly in the industrial Midwest and Northeast, are well-known for high levels of racial segregation. This is especially true where core cities end and suburbs begin; often crossing the street can lead to physically similar, but much less ethnically diverse, suburban neighborhood. While these differences are often visually or “intuitively” apparent, this study seeks to quantify them using Geographic Information Systems and a variety of statistical methods. 2016 Census block group data are used to calculate an ethnic Herfindahl index for a set of two dozen large U.S. cities and their contiguous suburbs. Then, a mathematical method is developed to calculate a block-group-level “Border Disparity Index” (BDI), which is shown to vary by MSA and by specific suburbs. Its values can be compared across the sample to examine which cities are more likely to have borders that separate more-diverse block groups from less-diverse ones. The index can also be used to see which core cities are relatively more or less diverse than their suburbs, and which individual suburbs have the largest disparities vis-à-vis their core city. Atlanta and Detroit have particularly diverse suburbs, while Milwaukee’s are not. Regression analysis shows that income differences and suburban shares of Black residents play significant roles in explaining variation across suburbs.

\end{abstract}

\section{Introduction}
Many U.S. metropolitan areas, such as Detroit and Philadelphia, are well-known for racial segregation. In many places, there are stark differences where the core city and its suburbs meet; often crossing a single street can lead one into a neighborhood with a much higher degree of diversity. These adjoining neighborhoods often appear physically similar, and while one might assume that economic forces should lead to similar populations in similar neighborhoods, this is clearly not the case. Segregation is not always manifested in the stereotypical case of mostly white suburbs located across the street from a diverse city; oftentimes a whiter, yet diverse, suburb can be adjacent to a monoethnic (often Black) city neighborhood. Since leaving the segregated city might increase diversity, or entering a whiter suburb might decrease it, these disparities must be examined individually. 

These “dividing lines” are often known locally; Detroit’s northern border at 8 Mile Road has perhaps taken on the largest significance outside its immediate metropolitan area. Another, less well-known example is where the city of Milwaukee borders the eastern part of Wauwatosa at 60th Street: Even though the housing stock and overall urban environment are similar, suburban homes are more expensive, and the population is more likely to be white, on the suburban side of the border. Sometimes the situation is reversed, however; Chicago’s Austin neighborhood on its West Side is almost entirely African-American, so suburban Oak Park (across Austin Avenue) is more ethnically diverse. Similar examples can be found elsewhere in the country, but they are by no means universal.

Both geography and economics, as well as local policies, can drive variations in these “border disparities.” While older cities in the Northeast and Midwest have clearly delineated borders that do not encompass their entire urbanized areas, the “overbounded” cities in the South and West include areas that would be considered suburban or even rural elsewhere. City borders are often jagged or non-linear, often reflecting a history of annexation policy and municipal consolidation. In addition, the overall degree of diversity, both in the core city and in individual suburbs, will determine how stark any potential contrast might be. Other economic factors, such as relative income, might also help explain these contrasts. Finally, the length of common borders might help determine the degree of exposure among different types of neighborhood. 

While any disparities where cities meet their suburbs are often visually or “intuitively” apparent, this study seeks to quantify them using Geographic Information Systems and a variety of statistical methods. Census data for race at the block group level for 2016 are used, for a set of two dozen large cities and their surrounding suburbs. After calculating an ethnic Herfindahl index to measure racial diversity, spatial patterns in the index are examined, before a method is developed to capture sharp differences at the borders between the core city and each suburb. 

This approach calculates a block-group-level “Border Disparity Index” (BDI), for diversity; these indices vary by MSA and by specific suburban borders. The values of this index can be compared across the sample to examine which cities are more likely to have borders that separate more-diverse places from less-diverse ones. The index can also be used to see which MSAs have suburbs that are more or less diverse than the city side of the border, and to identify specific suburbs where these differences are largest. As noted above using “intuitive” methods, the city of Wauwatosa does indeed have one of the highest BDI values, as does Detroit’s northern suburb of Warren. An econometric estimation shows that, after controlling for the length of common borders and the overall diversity levels in each suburb compared to the core city, relative income differences and the percentage of Black suburban residents are shown to be significant determinants of variation in border disparities. The same technique is used to calculate disparities in block groups’ proportions of Black residents as well; the areas with the largest disparities are often very different. 

\section{Previous Literature}
The geographic patterns of segregation, as well as its underlying mechanisms, have been extensively studied in the literature. Oftentimes economic segregation is the focus, as in Jargowsky (1996), but this type of segregation is often studied in tandem with racial segregation (as in Lichter et al., 2012; or Hero and Levy, 2016). South et al. (2011) examine variations for nearly 9,000 households in 269 metropolitan areas, modeling tract-level racial composition as a function of a set of socioeconomic variables.

Other studies model inter-neighborhood mobility, or the lack thererof. South and Crowder (1997), for example, model the likelihood that poor residents are able to leave distressed neighborhoods; they note that this probability is lower for Black households. Crowder et al. (2012) also find that Black or White households are relatively unlikely to move into multiethnic neighborhoods. As a result, “dividing lines” between neighborhoods might be persistent.

A relatively large share of research on economic segregation is conducted across urban areas. Downey (2003) notes that ethnic segregation is often measured at the place, rather than the disaggregated, level.  Lichter et al. (2015) examine within- and between-place measures of racial segregation for 222 metropolitan areas, finding that segregation had declined since 1990, particularly at the “micro,” or neighborhood, level. They note that Chicago, followed by Milwaukee, has the highest level of “macro,” or place-level segregation, which is often increasing. Anacker et al. (2017), however, focus on diversity among suburbs, finding that “mature” suburbs have levels of segregation similar to those of central cities, while newer ones do not. Relatively little is done, however, to examine small geographic areas within cities and suburbs.

Key analyses of ethnic boundaries and discontinuities across space have identified neighborhood-level segregation, which is often measured with the well-known indices of isolation, exposure, clustering, and evenness; these are discussed in detail by Massey and Denton (1988). Logan et al. (2011) develop three methodological approaches in a study of the U.S. city of Newark in 1880: K-functions and clustering; an "energy-minimizing algorithm"; and a Bayesian approach. Siegel-Hawley (2013) analyzes the role school of school district boundaries in fostering segregation in four Southern U.S. cities. 

Much of this literature has focused on neighborhoods in England, however. Dean et al. (2018) highlight the effects of "social frontiers," which include weakened social networks and a loss of social control in these areas. Social frontiers are also shown to be associated with increased crime rates in Sheffield. The concept of spatial discontinuities has also been applied to English cities. Harris (2014) develops a mathematical method of identifying disparities in the proportions of White and Asian residents among small areas across the country; these differences are shown to have become smaller between 2001 and 2011. In an analysis that is highly relevant to the current study, Mitchell and Lee (2013) examine the role of physical features in creating discontinuities that can affect such relationships as spatial autocorrelation. An analysis of Glasgow shows that there is a weak association between these barriers and an index of socioeconomic deprivation. Water barriers such as rivers and canals, as well as open spaces, have more of an effect than do parks, railroads, highways, or other such features.  

In the United States, one study that calculates tract-level measures of neighborhood differences was conducted by Chakravorty (1996), whose “Neighborhood Disparity Index” (NDI) calculates each area’s value versus its surrounding areas’ average values. After comparing a number of variations and modifications of this measure, that study notes that core city NDI values often exceed their neighboring metro values. While that type of analysis is useful to capture city- or metro-wide variations, it does not capture the border effects that are the object of the current study. 

This analysis, therefore, modifies the NDI measure to focus on patterns on either side of a city border. By examining changes in a measure of neighborhood diversity when block groups across the city border are excluded, it is possible to isolate areas where these disparities are largest. It is then possible to make inferences for metropolitan areas as well as for core cities and for individual suburbs. This paper proceeds as follows. Section 3 describes the methodology. Section 4 explains the empirical results. Section 5 concludes.

\section{Methodology}
This study focuses specifically on disparities in ethnic diversity in Census block groups on both sides of city/suburban borders. For a set of 23 core cities (which includes 25 U.S. cities with populations above 250,000) and their immediately adjacent suburbs, Census data (2016 ACS 5-year estimates) are first used to calculate an ethnic Herfindahl index to measure diversity within each block group:   	

\begin{equation}
H =1 - \sum\limits_{i=1}^5 p_i^2
\end{equation}

Here, $p$ is the proportion of the population of one of five groups (White, Black, Asian, Latino, and Other). 

\subsection{Calculating the Border Disparity Index}
Using this index, a method is developed to calculate “border disparities” across the sample. This builds upon the above-mentioned “Neighborhood Disparity Index”  developed by Chakravorty (1996), who measured differences between the value for a specific spatial unit and the average values for its neighboring units, and  can be plotted across an entire city or metropolitan area. It captures block-group-level disparities anywhere in a city or region and isolates block groups that are significantly more or less diverse than their immediate surroundings. 

This index is modified for the current study to account specifically for a “border effect,” by calculating the difference between two measures of the NDI as shown below in Equation (2). The unadjusted version (NDI\textsuperscript{U}) averages the diversity scores for all neighbors, including those that lie on the other side of the city-suburban border. The adjusted version (NDI\textsuperscript{A}) omits “city” neighbors for all suburban block groups and “suburban” neighbors for all city block groups. The difference between the two measures represents the relative amount of diversity (positive or negative) across the border from the city to a suburb, or vice versa. This “Border Disparity Index” (BDI) can therefore isolate specific areas where these differences are largest, at the metro, place, or block-group level.

Mathematically, the NDI for a specific block group measures the average H index value of surrounding block groups; in this sense, it is equivalent to a spatial lag estimate. The unadjusted NDI\textsuperscript{U} is therefore calculated using a row-normalized Queen contiguity matrix of order one; this (n x n) weights matrix W\textsuperscript{U} is multiplied by the (n x 1) vector of H indices to create an (n x 1) vector.

Next, the weights matrix is adjusted to remove any city-suburban neighbors. This is done by coding a vector with the value of +1 for City block groups and -1 for Suburban block groups. The resulting (n x n) product matrix will have values of +1 for all city-city or suburban-suburban pairs; any entry a\textsubscript{ij} with a value of -1 in this matrix (which measures city-suburban pairs) will have its corresponding entry replaced by zero in the W\textsuperscript{A} matrix and the row re-normalized to sum to one. The resulting (n x 1) vector NDI\textsuperscript{A} therefore does not include cross-city-border neighbors. The corresponding equation for the index is as follows: 

\begin{equation}
BDI = \textbf{W}^UH - \textbf{W}^AH  = NDI^U - NDI^A
\end{equation}

Since all non-border BDI values will by definition be zero, these are omitted in this analysis. All that remains are the subset of block groups that lie on either side—both city and suburban—of the core city’s border. Figure 1 provides a visual example of a hypothetical high-diversity area (each block group’s H index equals 0.8), separated by a border with a low diversity area (H values equal 0.0). The chosen Area 1 block group has a BDI of -0.3, while the selected Area 2 block group’s BDI is +0.3. Obviously, the shapes and numbers of neighbors will vary, but it is clear that positive values occur when diversity is kept out, and negative values result when diversity is kept in.

\begin{figure}[ht]
\hfill
\caption{An Example of the Calculation of the Border Disparity Index.}
\begin{center}
\includegraphics[width=.3\textwidth]{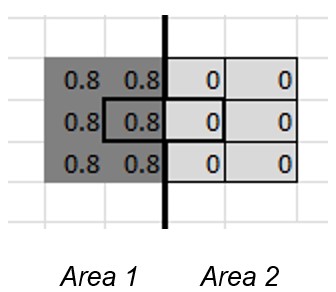}
\caption*{\textbf{Twelve hypothetical block groups:}\\
Central block groups in question: Outlined with black lines (Area 1 H = 0.8; Area 2 H = 0.0)\\
Border: Thick vertical line\\
\\
\textit{\textbf{Calculated values:}}\\
\textbf{Area 1 block group (H = 0.8)}\\
NDI\textsuperscript{U} = 4.0/8 = 0.5\\
NDI\textsuperscript{A} = 4.0/5 = 0.8\\
BDI = -0.3 (diversity kept in by border)\\
\\
\textbf{Area 2 block group (H = 0.0)}\\
NDI\textsuperscript{U} = 2.4/8 = 0.3\\
NDI\textsuperscript{A} = 0.0/5 = 0.0\\
BDI = +0.3 (diversity kept out by border)\
}
\end{center}
\end{figure}

While one might expect disparities to be large in the older, denser, underbounded cites of the Northeast and Midwest, this study attempts to include as large of a sample as is possible with the available data. The choice of cities requires that they satisfy three criteria. First, they must be large enough to have a sufficient number of block groups to analyze, both within the city and in their surrounding suburbs. For this reason, potential candidate cities are restricted to those with populations above 250,000. Also excluded are cities, particularly in the South and West, with the majority of their populations in the core city and large, sparsely populated block groups surrounding it. Second, an attempt is made to focus on cities with reasonably straight borders. Because of annexation and other political factors, some cities have “jagged” shapes. These are evaluated visually by the author. Third, while internal suburbs (such as Detroit-Hamtramck) are allowed, cities that have “exclaves” that are entirely separate from the main part of the city are not. 

This leaves a final set of 23 cities and their suburbs. For two of these (Minneapolis-St. Paul and Denver-Aurora), two large cities are included as a single entity. In other cases, all cities of any size are treated as suburbs. Long Beach, for example, is considered to be a suburb of Los Angeles even though it is larger than many of the core cities examined here. Suburbs are defined as all Census-designated places that touch the core city. Not every point on the core city boundary is abutted by a Census-designated place, however. While in some cases (for example, Michigan and Pennsylvania) state data were available for townships that were not considered to be Census “places,” this was not the case for every state. Suburbs across state lines (such as Chicago-Hammond, IN) were included, but not those located across rivers or other bodies of water. In all, a total of 357 suburbs are included in this study.

\subsection{Metro- and place-level analysis}
For each city or suburban block group that is located on a city border, BDI is then calculated, which can be used for global, metro-level, or place-level analysis. First, the H and BDI indices are examined, to calculate summary statistics that can be used for statistical inference. Because this index is newly created, one cannot assume that traditional statistics automatically apply. Particular attention is paid to the 5\% and 95\% quantiles of the distributions. These thresholds vary by metro, and provide useful information regarding where border disparities are highest. Metros are also identified where the maximum BDI value is located in the core city; this might indicate that suburbs could be relatively more diverse than the core city. Block groups with BDI values greater than two standard deviations from the mean (in absolute value) are also identified, and the proportion of each metro’s “city” or “suburban” block groups that exceed this threshold are calculated. 

Specific suburbs are then examined at the place level. With more than 350 suburbs, a metric is first established to isolate the places with the largest disparities. These include the maximum and minimum BDI values, as well as the range, the sum, and the mean value. Because varying lengths of common borders, as well as other factors, need to be accounted for, the maximum BDI value within a place is chosen as a proxy for its overall degree of disparity. This does not minimize any potential limitations with this choice. Using this measure, as well as a rough 5 percent cutoff, the 18 suburbs with the largest maxima and the 18 with the smallest minima are identified as well. 

\subsection{Regression analysis}
Finally, variation in each suburb’s maximum BDI value is modeled as a function of certain geographic and socioeconomic variables. These include:
\textit{H} = Each suburb’s diversity index, and its difference (\textit{HGAP}) with the core city. 
The length of the common core-city/suburban \textit{BORDER}, and the percent of the total (\textit{PERCBORDER}). Longer borders provide more observations, and therefore more opportunities for extreme values.
The percentage of Black residents (\textit{PERCBLK}) and the differences with the core city (\textit{BLKDIFF}). The difference in percentages of White residents (\textit{WHTDIFF}) is included as well. It is expected that, because of past and present patterns of racial discrimination, that disparities might be particularly high where the percentage of Black residents is high. 
Each suburb’s median income (\textit{MEDINC}) and its ratio versus the core city value (\textit{MEDINCRAT}). Equal values will by definition equal one. It is expected that disparities might be higher where income differentials are largest.
Each suburb’s population density (\textit{POPDENS}) and its ratio versus the core city (\textit{POPDENSRAT}). The relative size of the suburb compared to the core city is captured by \textit{POPRATIO}.

Dummies for various regions (such as the Northeast, South, “Rust Belt,” etc.) were included in additional test specifications, but were not significant and were therefore not added to the model. This model is be estimated using Ordinary Least Squares, with standard errors clustered by MSA. Various specifications will be evaluated side-by-side, with variable significance and adjusted R-squared evaluated when selecting the best model.

Much of this analysis is repeated, using block groups’ percentage Black in place of the H index. The results differ widely, at the MSA, place, and block-group level, and the determinants shown in the regression model vary as well. While more research must be done to examine the consequences of segregation uncovered here, these results highlight stark differences between overall diversity measures and this proportion. The main results are presented below.

\section{Results}
\subsection{Calculating and testing the index}
Using the formula mentioned above, the H index is calculated for each of the metro areas in this study. As a preliminary test, high- and low-value clusters are also calculated for the Milwaukee area using a more traditional method: The local Moran’s I statistic available in the GeoDa software package. When the Border Disparity Index introduced in this study is calculated, it can be compared against this method to see whether it captures the same spatial patterns. The leftmost map of Figure 2 presents the distribution of the H index across the Milwaukee area; high-diversity block groups are visible on the far Northwest side, for example, right up to the city border. 

\begin{figure}[ht]
\hfill
\caption{Diversity Indices, Clusters, and Border Disparities in Milwaukee, 2016.}
\includegraphics[width=.34\textwidth]{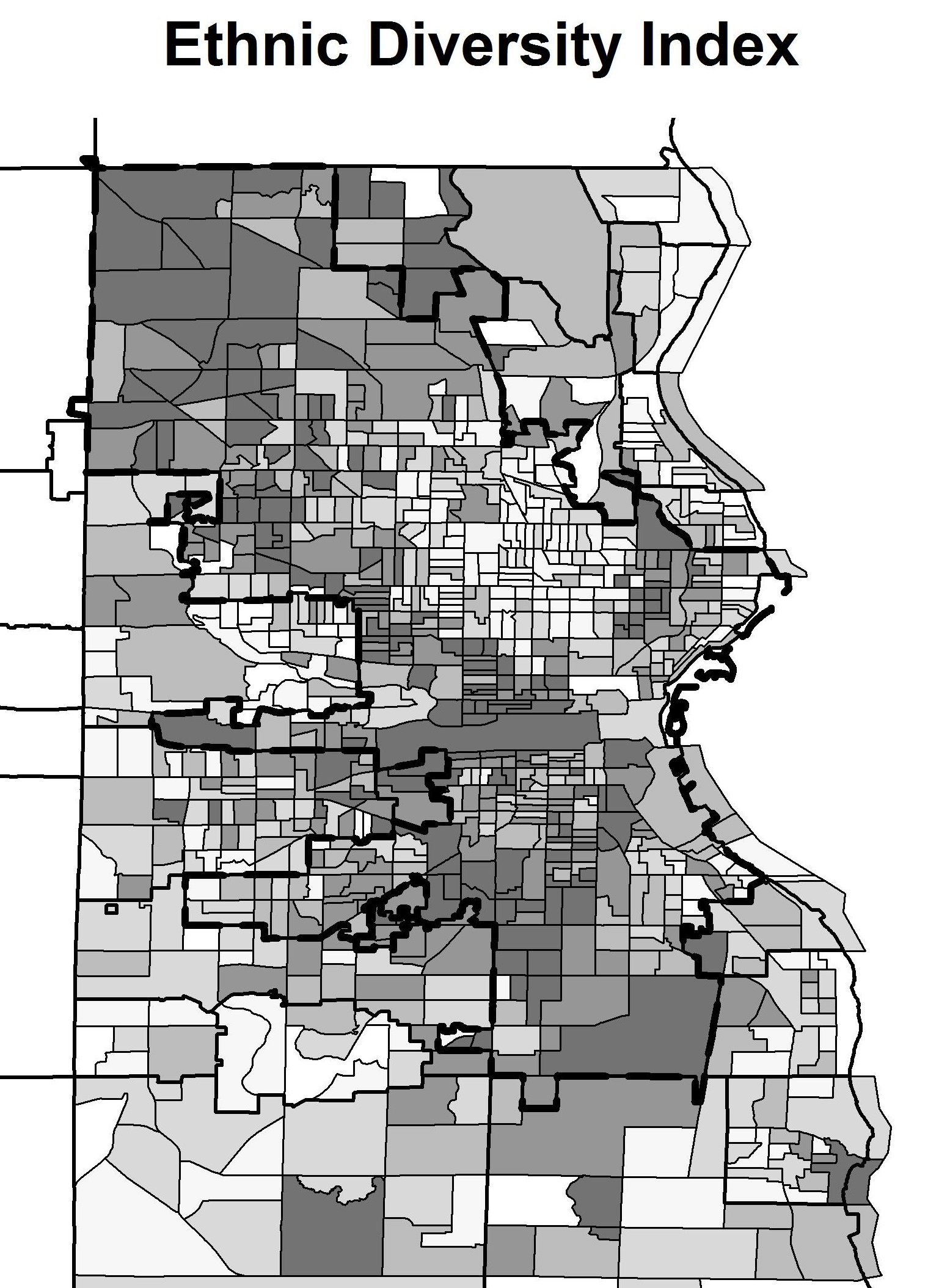}
\includegraphics[width=.27\textwidth]{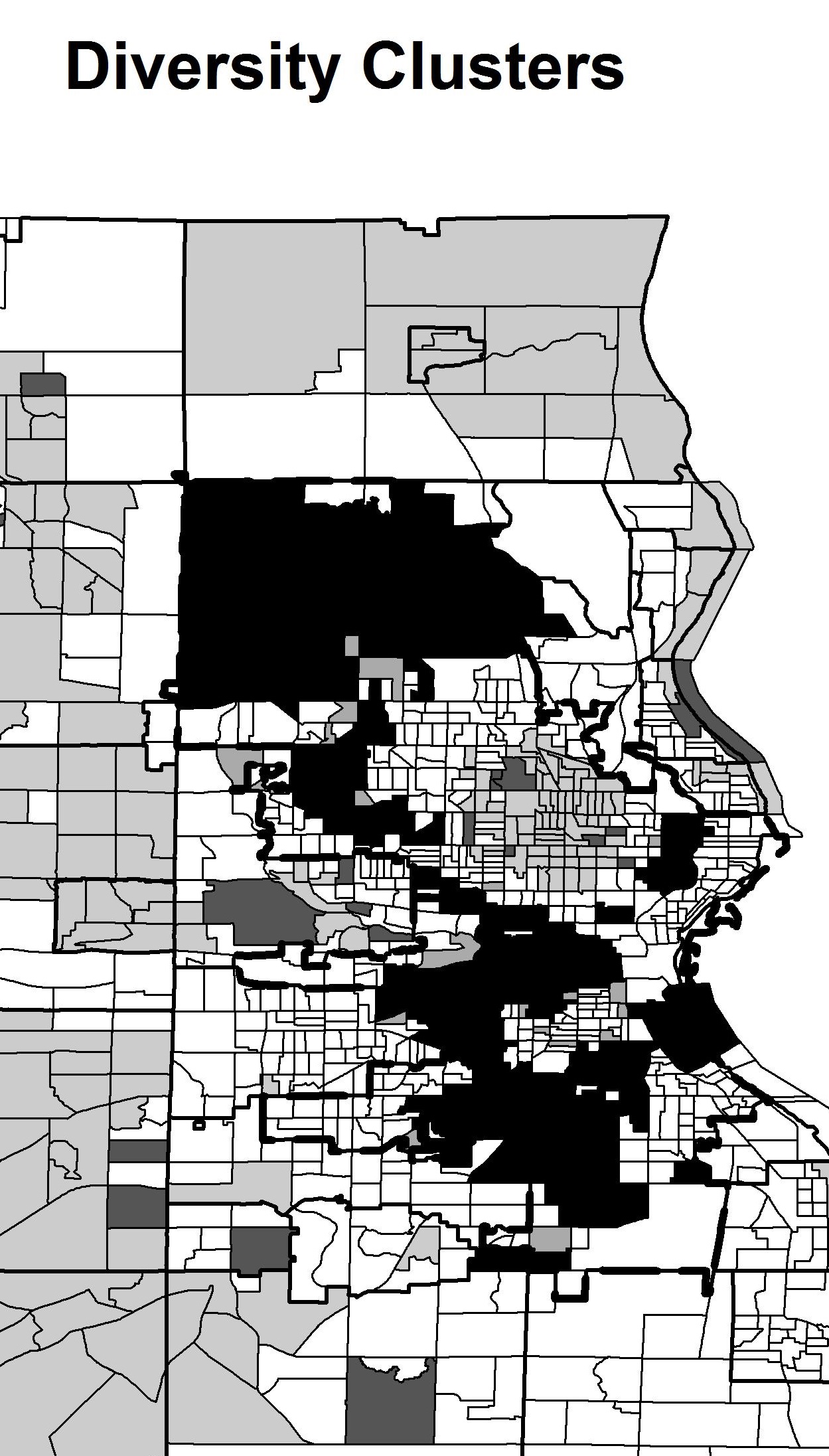}
\includegraphics[width=.36\textwidth]{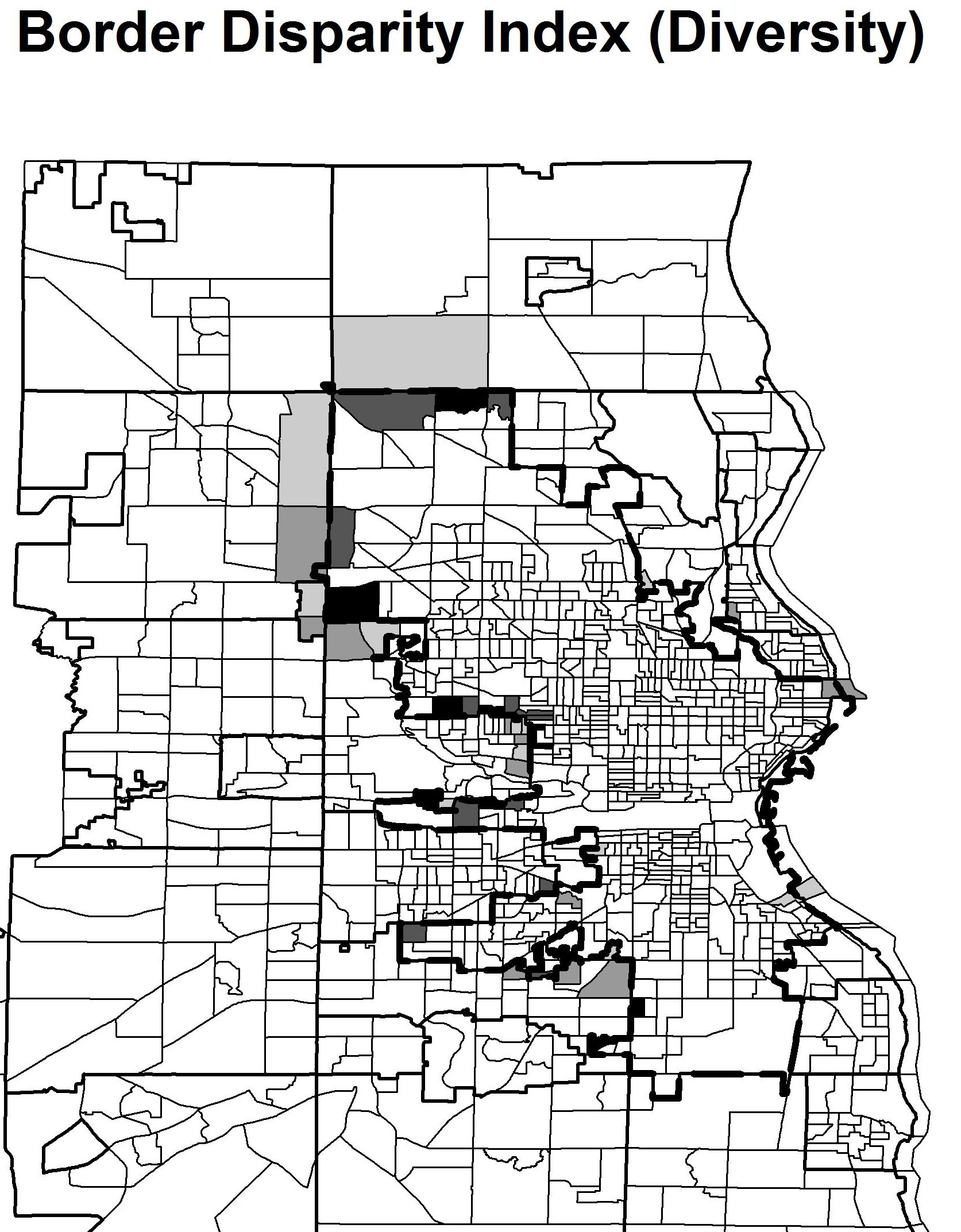}
\caption*{Diversity calculated as H-index. Darker = higher value.\\
Clusters calculated using the univariate local Moran’s I statistic in the GeoDa software. Black = “High-High.”\\
Border Disparity Indices: (Dark grey/black = high, light grey = low, white = not calculated)
}

\end{figure}

\begin{table}[ht]
\caption{Summary Statistics of the Diversity (Herfindahl) and Border Disparity Indices.}
\begin{center}
\begin{tabular}{lrrrrlrrrr}

&H&BDI(H)&BDI(\% Black)\\
\hline
Mean&0.368&0.002&-0.000\\
St. Dev&0.206&0.061&0.077\\
Min&0.000&-0.349&-0.493\\
1Q&0.194&-0.030&-0.018\\
Median&0.382&0.003&-0.001\\
3Q&0.543&0.034&0.016\\
Max&0.795&0.369&0.649\\
N&33484&4544&4544\\

\hline
\end{tabular}
\end{center}

\end{table}

High H values can also be found in “outer city” areas between the traditional inner city and certain close-in suburbs. The map of the clusters (“high-high”) capture these trends. While “high-low” areas do not lie immediately across the city border, there is a “buffer zone” in those areas, particularly in the eastern part of Wauwatosa. The BDI reflects these same patterns. In particular, high values can be found on Milwaukee’s border with Wauwatosa, as well as further north. 

Table 1 provides summary statistics for the H-index for all 33,484 block groups in the sample, as well as for the Border Disparity Index and the percentage of Black residents for the 4,544 block groups that lie on either side of a city border. As might be expected, “diversity” not equivalent to “percent Black”; this highlights differences in policy, which often treats the two concepts as synonymous in the academic literature. While not presented in as great of detail here, conducting the analysis for the percentage Black instead of the H index highlights how different the two measures are.

Both Border Disparity Indices are fairly normally distributed, but the one for the H index is less skewed. It ranges from -0.349 to +0.369, with a mean of 0.002 and a standard deviation of 0.061. The BDI for the percentage Black ranges from -0.493 to +0.649, with a mean of almost zero and a standard deviation of 0.077. In addition to being more skewed, the second measure also has higher kurtosis, with a larger percentage (roughly 4.75 vs. 2.75) of block group values greater than two standard deviations from the mean. These statistics will be used when comparing “extreme” disparities for both metro areas and individual places.

\subsection{Metro-level analysis}

	Next, each of the 23 metro areas are compared against one another. Because the most diverse cities might have the largest disparities, each city is ranked by its median H index in Figure 3.  Los Angeles is most diverse on average, and Detroit and Buffalo the least diverse. The spread of 95 percent values and 5 percent values is wide overall, but appears to be fairly uniform across metros. One exception is that Minneapolis-St. Paul has a relatively high 95\% value, while Milwaukee and Portland have “compressed” distributions with high 5\% and low 95\% values. This suggests that these entire regions’ suburbs are relatively homogenous.


\begin{figure}[ht]
\hfill
\caption{Metro Diversity Indices for All Block Groups, Sorted by Median Value.}
\begin{center}
\includegraphics[width=.5\textwidth]{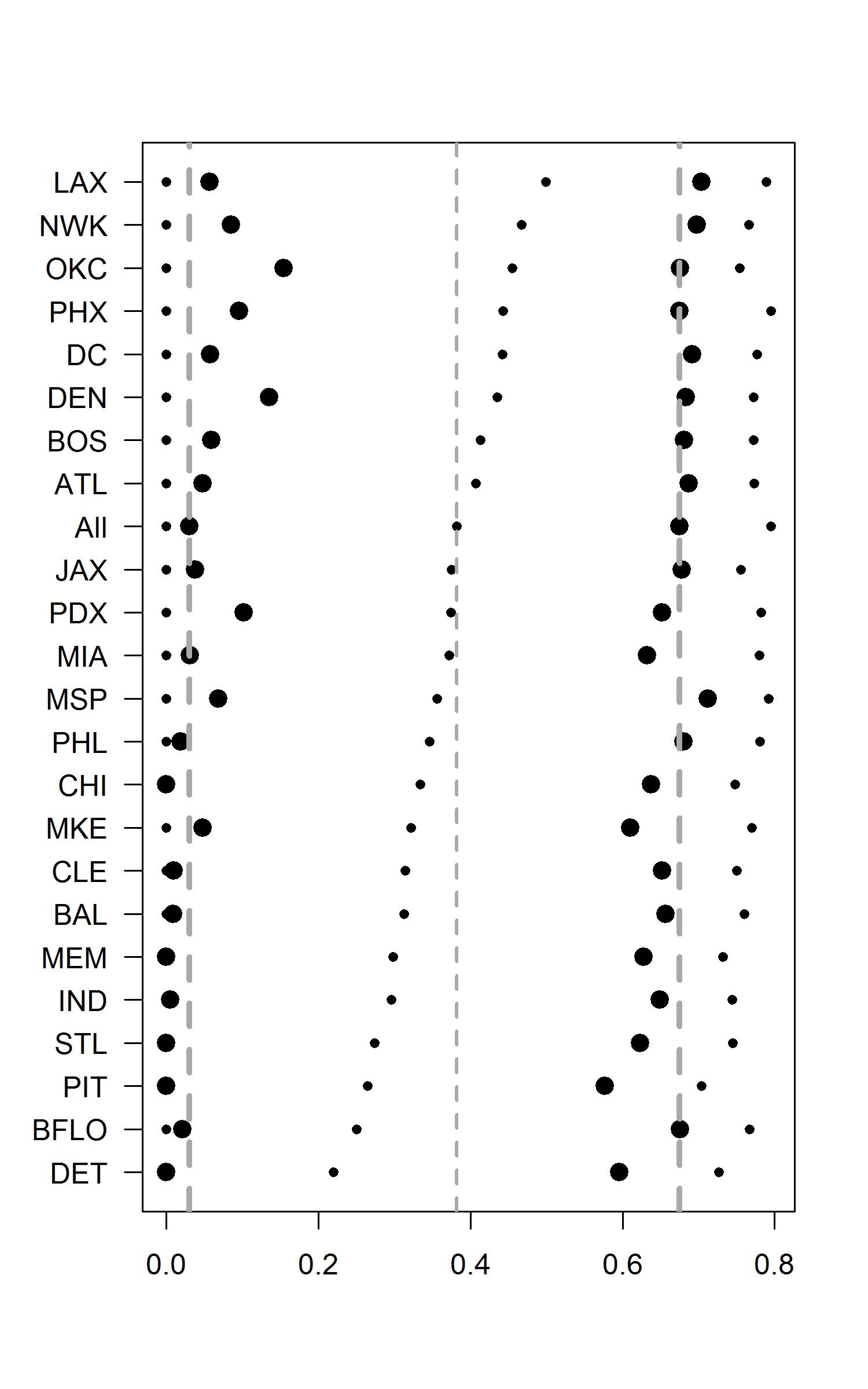}
\caption*{Horizontal lines indicate full-sample 5\%, median, and 95\% values.\\
Circles represent each MSA's minimum, 5\%, median, 95\%, and maximum values.}
\end{center}
\end{figure}

\begin{table}[ht]
\caption{Selected Statistics for Core Cities and Their BDI (H-Index) Statistics.}
 \begin{center}
\begin{tabular}{llrll}

MSA&City&\# Suburbs&City Max&City Min\\
\hline
ATL&Atlanta&13&0.285*&-0.073\\
BAL&Baltimore&11&0.210*&-0.102\\
BFLO&Buffalo&7&0.176 &-0.136\\
BOS&Boston&10&0.204*&-0.169\\
CHI&Chicago&35&0.369*&-0.235\\
CLE&Cleveland&13&0.175&-0.137\\
DC&Washington&16&0.202*&-0.074\\
DEN&Denver-Aurora&18&0.143&-0.151\\
DET&Detroit&20&0.205*&-0.222\\
IND&Indianapolis&15&0.084&-0.166*\\
JAX&Jacksonville&10&0.037&-0.163*\\
LAX&Los Angeles&35&0.239*&-0.300*\\
MEM&Memphis&8&0.144&-0.224*\\
MIA&Miami&10&0.140*&-0.061\\
MKE&Milwaukee&14&0.106&-0.286*\\
MSP&Minneapolis-St. Paul&12&0.176*&-0.146\\
NWK&Newark&3&0.116&-0.130\\
OKC&Oklahoma City&23&0.092&-0.230*\\
PDX&Portland&10&0.082&-0.099\\
PHL&Philadelphia&20&0.289*&-0.087\\
PHX&Phoenix&12&0.108&-0.135\\
PIT&Pittsburgh&28&0.152&-0.349*\\
STL&St. Louis&14&0.125*&-0.138\\
\hline
\end{tabular}
\end{center}
\caption*{* = Maximum or minimum region-wide value located inside core city.}
\end{table}

Table 2 provides a summary of statistics related to core cities and their environs; Chicago has the most suburbs, as well as the highest in-city BDI value. This suggests that diversity is kept “out” of the city by at least part of its border. The largest BDI value is in the core city rather than in a suburb for 12 of the 23 metros, including Atlanta, Detroit, Washington,  Cleveland, and Miami. Nine metros, including Pittsburgh, Milwaukee, Jacksonville, Memphis, and Indianapolis, have their smallest BDI in the core city, indicating that diversity is kept “inside” the core cities. Los Angeles has both the largest and the smallest values within the city itself, suggesting that both effects can be present along different sections of the city border. 

\begin{figure}[ht]
\hfill
\caption{Proportion of MSA Block Groups  $\pm2$ Standard Deviations from Full-sample Mean.}
\begin{adjustwidth}{-.75in}{-.75in}  
\includegraphics[width=.65\textwidth]{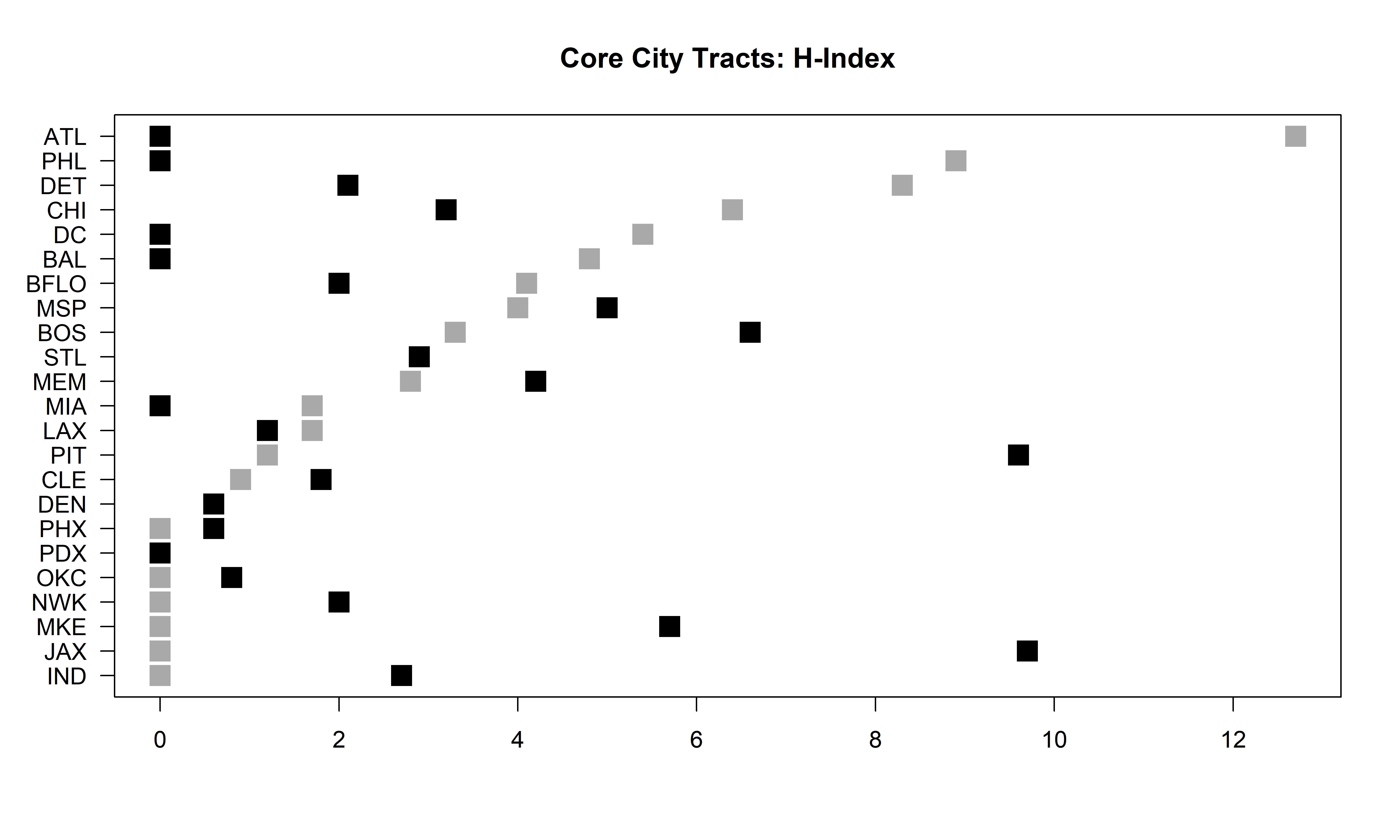}
\includegraphics[width=.65\textwidth]{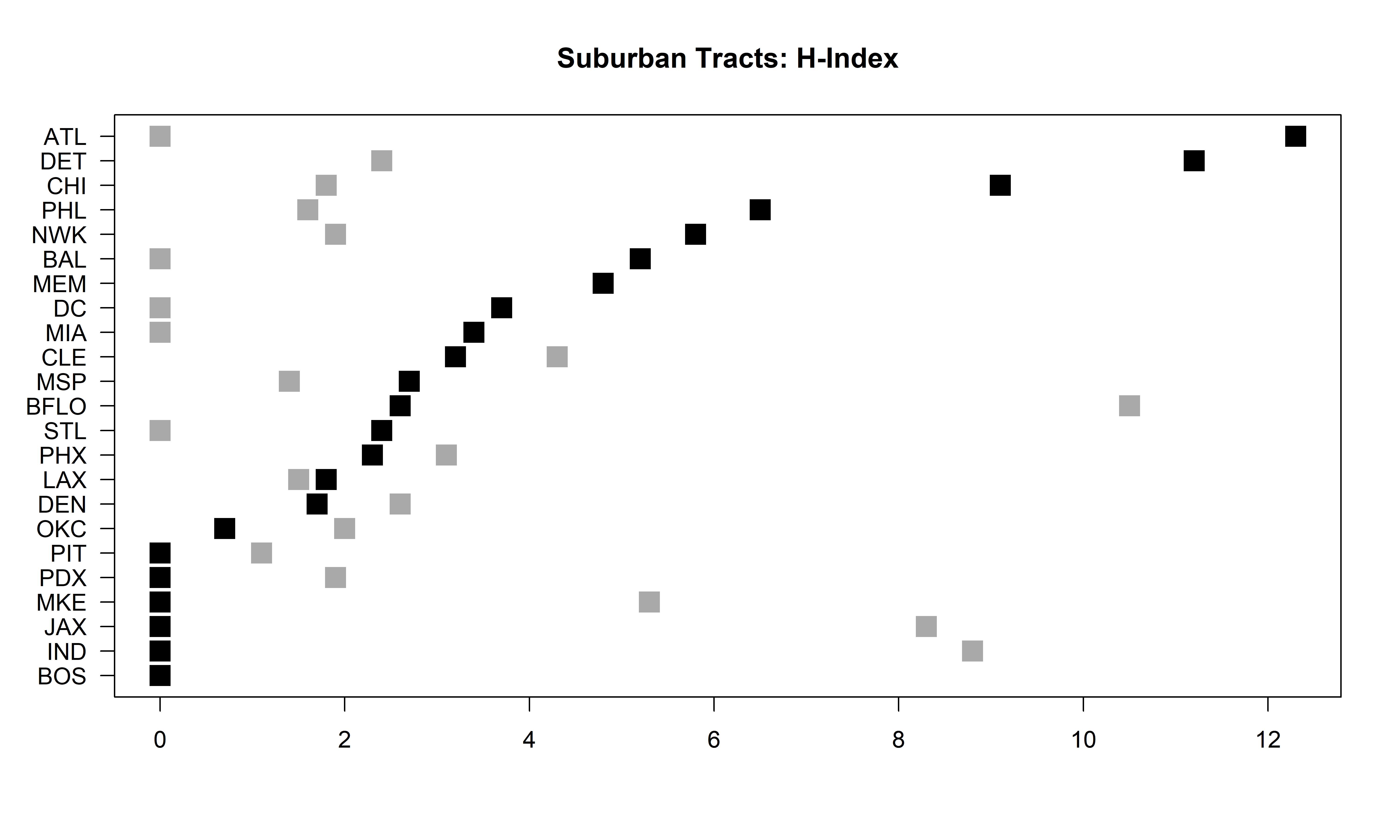}
\end{adjustwidth}
\caption*{Black squares = significantly large positive values (given city/suburban area more diverse than suburban/city neighbors)\\
Grey squares = significantly large negative values (given city/suburban area less diverse than suburban/city neighbors)}
\begin{adjustwidth}{-.75in}{-.75in}  
\includegraphics[width=.65\textwidth]{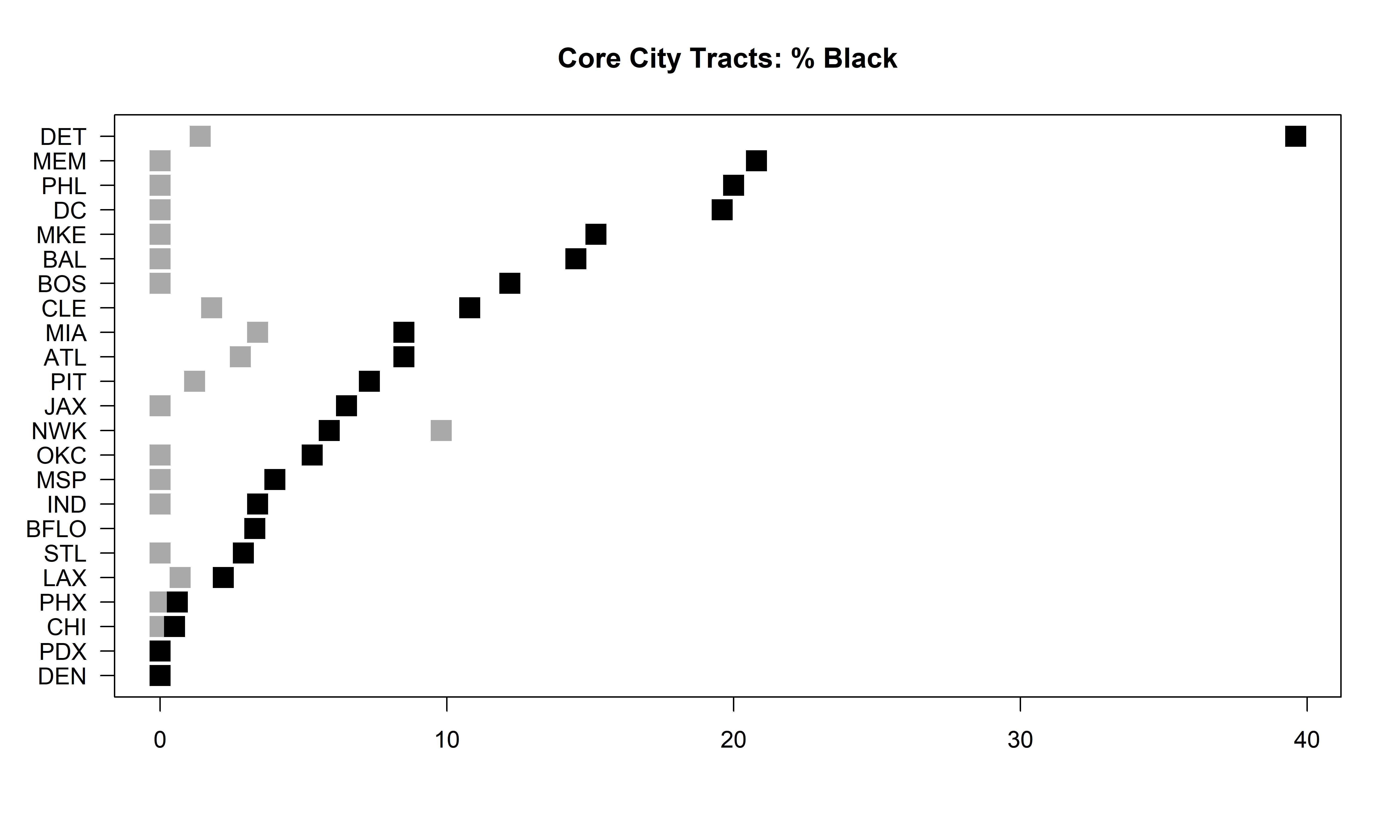}
\includegraphics[width=.65\textwidth]{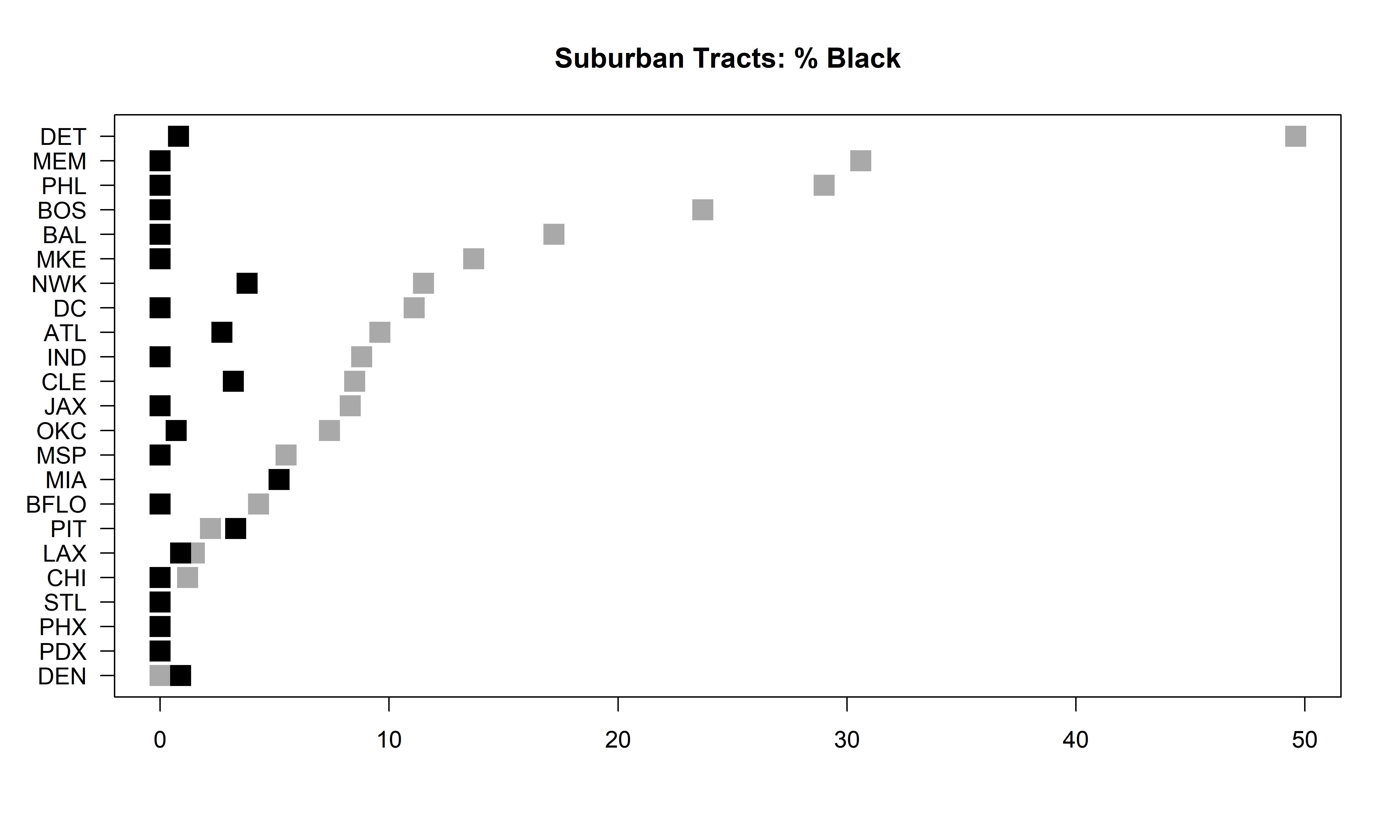}
\end{adjustwidth}
\caption*{Black squares = significantly large positive values (given area has larger Black population than neighbors)\\
Grey squares = significantly large negative values (given area has smaller Black population than neighbors)}
\end{figure}

Metros with large border disparities can also be identified as those with large proportions of “extreme” BDI values. Figure 4 presents the proportion of block groups in each metro’s core city and suburbs with BDI values more than two standard deviations from the full-sample mean. These proportions are generally presented in decreasing order of significantly negative city, or positive suburban, shares. By far, Atlanta has the largest share of extreme negative suburban shares and extreme positive city shares; this suggests that city neighborhoods are likely to be less diverse along the border, while adjacent suburbs are likely to be more diverse. The same can be said, to a lesser extent, for Philadelphia, Detroit, and Chicago.	While Atlanta also has a very low proportion of city block groups that keep diversity in, and of suburban borders that keep diversity out, the opposite situation is shown for Indianapolis, Jacksonville, and Milwaukee. These areas have large proportions of city block groups with high BDI values and suburban block groups with low BDI values. They also have small proportions of city block groups with low BDI values and suburban block groups with high BDI values. In addition, Buffalo’s suburbs have a very large share of block groups with significant negative values. This corresponds to these core cities being relatively diverse relative to their suburbs. Chicago, on the other hand, has city neighborhoods with both high and low BDI values, highlighting variation across its vast metropolitan region. On the other extreme, Portland has few extreme BDI values for either city type.

A similar analysis for the BDI index that captures disparities in the percentage Black is different in one key aspect: Most MSAs have few significantly negative city BDI values or positive suburban values. Most cities have larger shares of Black residents inside the border.  Detroit has the largest racial disparities, followed by Memphis and Philadelphia. Milwaukee and Baltimore are close behind. Again, “diversity” is not synonymous with “percentage Black.”

\subsection{Place-level analysis}

Which specific suburbs within this sample of metros show the largest border disparities? To answer this question, one first must choose an appropriate aggregate measure that incorporates all relevant block groups within each suburb. Five alternatives are evaluated before proceeding. Mean BDI would incorporate the entire border, but might penalize situations in which both extremely high values and extremely low values are simultaneously present. The sum of all BDI values within a suburb would do the same, as well as penalize single, extremely high values in favor of “long borders” that have a large number of relatively low values. After showing correlations among the measures, the maximum and minimum BDI value within each suburb are chosen to serve as a measure of the whole. One must be aware of potential problems, particularly cases in which a single outlier, perhaps unique to that location, might drive an entire city’s results, however. Table 3 shows the Spearman correlations, as well as with the range of values within the suburbs; the mean and maximum values are most correlated. The maximum value is therefore shown to be an adequate measure of border disparities at the 
place level. 

\begin{table}[ht]
\caption{Correlations Among BDI Index Values (H-Index, N = 379 Places).}
\begin{center}
\begin{tabular}{lrrrrr}

&Mean&Sum&Max&Min&Range\\
\hline
Mean&1&0.706&0.795&0.829&-0.058\\
Sum&&1&0.660&0.645&-0.007\\
Max&&&1&0.420&0.516\\
Min&&&&1&-0.561\\
Range&&&&&1\\
\hline
\end{tabular}
\end{center}
\end{table}

Of the 357 suburbs in this study, the five percent of suburbs with the highest maximum BDI\textsuperscript{H} values are listed, as well as the five percent with the lowest minimum values, in Table 4. The first 18 suburbs are those that are considerably less diverse than their nearby city neighborhoods. The suburb with the largest BDI value is Hometown, Illinois, on the southern border of Chicago. In second place is Wauwatosa, Wisconsin, which was highlighted earlier in this study. Dearborn, Michigan, just south of Detroit, appears near the middle of the list. Interestingly, four of these 18 suburbs are in the Indianapolis area. The 18 suburbs with the lowest minimum values, which are themselves more diverse than neighboring cities' areas, include Cheltenham and Springfield, outside Philadelphia; Cheektowaga, New York (just east of Buffalo’s city line); Cicero, Illinois; and Warren, Michigan. Four of these 18 suburbs border Detroit.

\begin{table}[ht]
\caption{“Extreme” (5\%) Maximum or Minimum BDI Values (357 Suburbs) }
\begin{center}
\begin{tabular}{llrrrrr}

\multicolumn{7}{l}{\textit{H-index: Highest Maximum BDI Values (Suburb less diverse than core city)}}\\
\hline
MSA&Suburb&Mean&Sum&\textbf{Max}&Min&Range\\
\hline
CHI&Hometown&0.125&0.627&0.288&0.044&0.244\\
MKE&Wauwatosa&0.056&1.685&0.270&-0.029&0.299\\
IND&Cumberland&0.126&0.504&0.262&0.035&0.227\\
PIT&Reserve&0.109&0.544&0.260&0.007&0.253\\
PHX&Scottsdale&0.025&0.706&0.248&-0.034&0.282\\
LAX&Florence-Graham&0.039&0.665&0.232&-0.012&0.244\\
MEM&Germantown&0.067&0.943&0.231&-0.008&0.239\\
DEN&Holly Hills&0.144&0.287&0.208&0.079&0.129\\
DEN&Bow Mar&0.097&0.291&0.200&0.027&0.173\\
LAX&Glendale&0.010&0.200&0.198&-0.093&0.291\\
PHX&Tolleson&0.052&0.208&0.189&-0.013&0.202\\
IND&Fishers&0.022&0.134&0.188&-0.030&0.218\\
DET&Dearborn&0.048&0.667&0.187&-0.041&0.228\\
JAX&Orange Park&0.075&0.301&0.186&-0.002&0.188\\
IND&Southport&0.055&0.111&0.178&-0.067&0.245\\
IND&Speedway&0.087&0.695&0.176&0.017&0.159\\
CLE&Newburgh Heights&0.125&0.250&0.175&0.076&0.099\\
LAX&Inglewood&0.024&0.496&0.173&-0.069&0.242\\
CLE&Brooklyn&0.074&0.517&0.165&-0.010&0.175\\
\hline
\\

\multicolumn{7}{l}{\textit{H-index: Lowest Minimum BDI Values (Suburb more diverse than core city)}}\\
\hline
MSA&Suburb&Mean&Sum&Max&\textbf{Min}&Range\\
\hline
PHL&Cheltenham&-0.110&-1.648&-0.018&-0.320&0.302\\
PHL&Springfield&-0.067&-0.471&0.034&-0.320&0.354\\
DET&Warren&-0.123&-1.594&-0.033&-0.280&0.247\\
CHI&Cicero&-0.056&-0.670&0.033&-0.246&0.279\\
BFLO&Cheektowaga&-0.053&-0.527&0.029&-0.238&0.267\\
BAL&Rosedale&-0.137&-0.684&-0.079&-0.237&0.158\\
DET&Royal Oak Township&-0.146&-0.438&-0.085&-0.232&0.147\\
DET&Harper Woods&-0.079&-0.636&0.057&-0.203&0.260\\
LAX&South Pasadena&-0.078&-0.389&-0.039&-0.199&0.160\\
ATL&Hapeville&-0.113&-0.566&-0.043&-0.193&0.150\\
MEM&Horn Lake&-0.138&-0.414&-0.077&-0.192&0.115\\
DEN&Glendale&-0.147&-0.440&-0.064&-0.188&0.124\\
LAX&Burbank&-0.032&-0.389&0.056&-0.188&0.244\\
LAX&Alhambra&-0.067&-0.401&0.005&-0.188&0.193\\
ATL&East Point&-0.074&-0.959&-0.002&-0.177&0.175\\
CHI&Merrionette Park&-0.138&-0.275&-0.104&-0.171&0.067\\
CHI&Blue Island&-0.092&-0.644&-0.037&-0.171&0.134\\
PHX&Tempe&0.001&0.016&0.070&-0.167&0.237\\
DET&Southfield&-0.054&-0.430&0.015&-0.165&0.180\\
\hline
\end{tabular}
\end{center}
\end{table}

Table 5 also shows similar ranked suburbs for border disparities in suburbs’ percentage Black. While these need to be further investigated (including the lowest minimum value, located in a very small suburb within Oklahoma City), it is interesting that four Detroit suburbs are among the 18 suburbs with the highest BDI values (which would be relatively less Black than Detroit proper). Suburbs of Los Angeles dominate the 18 places with low minimum BDI values. At the same time, the Los Angeles area also has three suburbs with high BDI values. Three Chicago suburbs are among those with the largest BDI values, while three Pittsburgh suburbs have low minimum values. These findings may be worthy of a separate study.

\begin{table}[ht]
\caption{“Extreme” (5\%) Maximum or Minimum BDI Values (357 Suburbs) }
\begin{center}
\begin{tabular}{llrrrrr}

\multicolumn{7}{l}{\textit{Percent Black: Highest Max. BDI Values (Suburb lower percentage than core city)}}\\
\hline
MSA&Suburb&Mean&Sum&\textbf{Max}&Min&Range\\
\hline
DET&Grosse Pointe Park&0.301&2.106&0.649&0.189&0.460\\
MKE&Glendale&0.217&1.955&0.503&0.018&0.485\\
DET&Dearborn&0.220&3.086&0.462&-0.001&0.463\\
PHL&Lower Merion&0.123&1.601&0.446&0.013&0.433\\
DET&Dearborn Heights&0.169&1.014&0.387&0.075&0.312\\
MEM&Germantown&0.085&1.189&0.381&0.012&0.369\\
BUF&Cheektowaga&0.148&1.484&0.374&-0.037&0.411\\
MKE&Wauwatosa&0.050&1.500&0.362&-0.036&0.398\\
DET&Ferndale&0.226&0.904&0.358&0.144&0.214\\
OKC&Nichols Hills&0.112&0.560&0.334&0.004&0.330\\
DET&Warren&0.185&2.402&0.333&0.072&0.261\\
LAX&Inglewood&0.036&0.748&0.311&-0.095&0.406\\
MEM&Southaven&0.221&1.768&0.310&0.113&0.197\\
OKC&Jones&0.165&0.494&0.304&0.092&0.212\\
IND&Cumberland&0.174&0.694&0.300&0.092&0.208\\
PHL&Cheltenham&0.120&1.802&0.293&-0.086&0.379\\
PHL&Springfield&0.143&1.002&0.293&-0.003&0.296\\
DET&Redford Township&0.116&1.389&0.292&0.028&0.264\\
MEM&Bartlett&0.105&1.256&0.290&0.014&0.276\\
\hline
\\

\multicolumn{7}{l}{\textit{Percent Black: Lowest Min. BDI Values (Suburb Higher percentage than core city)}}\\
\hline
MSA&Suburb&Mean&Sum&Max&\textbf{Min}&Range\\
\hline
OKC&Forest Park&-0.127&-0.254&0.096&-0.350&0.446\\
NWK&East Orange&-0.054&-0.758&0.028&-0.298&0.326\\
CLE&East Cleveland&-0.045&-0.543&0.030&-0.218&0.248\\
DET&River Rouge&-0.041&-0.124&0.061&-0.200&0.261\\
LAX&Culver City&0.007&0.144&0.170&-0.198&0.368\\
LAX&Ladera Heights&-0.062&-0.312&0.039&-0.198&0.237\\
MIA&Brownsville&-0.077&-0.618&0.007&-0.188&0.195\\
PIT&Wilkinsburg&-0.024&-0.216&0.033&-0.133&0.166\\
LAX&West Rancho Dominguez&-0.058&-0.404&-0.001&-0.129&0.128\\
PIT&Homestead&-0.127&-0.253&-0.124&-0.129&0.005\\
BAL&Dundalk&-0.022&-0.198&0.020&-0.118&0.138\\
IND&Lawrence&0.058&0.864&0.249&-0.108&0.357\\
PIT&Mount Oliver&0.086&0.345&0.278&-0.098&0.376\\
LAX&Carson&-0.033&-0.198&0.000&-0.098&0.098\\
LAX&Inglewood&0.036&0.748&0.311&-0.095&0.406\\
LAX&View Park-Windsor Hills&-0.005&-0.047&0.042&-0.088&0.130\\
PHL&Cheltenham&0.120&1.802&0.293&-0.086&0.379\\
LAX&Westmont&0.015&0.230&0.187&-0.083&0.270\\
LAX&Gardena&-0.029&-0.407&0.002&-0.078&0.080\\
\hline

\end{tabular}
\end{center}
\end{table}

\subsection{Regression analysis}
	It is clear that these differences are driven by the economic, social, and geographic characteristics of each metro, core city, and suburb. Indianapolis, for example, takes up nearly an entire county and therefore has a much different urban structure than Detroit. To capture these differences, the results of a regression model, for the 310 suburbs for which full data were available, are provided in Table 6. Three specifications are given; adjusted R-squared rises as insignificant variables are removed.

\begin{table}[ht]
\begin{adjustwidth}{-1in}{1in}  
\caption{Regression Results, Suburban Maximum BDI Values.}
\begin{center}
\begin{tabular}{lrrrrrr}

&\multicolumn{3}{l}{DV = Max. BDI(H)}&\multicolumn{3}{l}{DV = Max. BDI(Percent Black)}\\
Variable&Coeff. (p-val.)&Coeff. (p-val.)&Coeff. (p-val.)&Coeff. (p-val.)&Coeff. (p-val.)&Coeff. (p-val.)\\
\hline
INPT&0.024 (0.495)&0.039 (0.171)&0.0200 (0.326)&0.028 (0.634)&0.014 (0.804)&-0.017 (0.400)\\
H&-0.063 (0.116)&-0.061 (0.110)&-0.0445 (0.170)&-0.026 (0.667)&&\\
HGAP&-0.061 (0.064)&\emph{-0.063 (0.035)}&\emph{-0.0801 (0.001)}&0.086 (0.150)&0.065 (0.052)&\emph{0.067 (0.019)}\\
BORDER&\emph{0.000 (0.001)}&\emph{0.000 (0.000)}&\emph{0.000 (0.000)}&0.000 (0.389)&0.000 (0.340)&\\
PERCBORDER&0.000 (0.073)&\emph{0.000 (0.017)}&\emph{0.0004 (0.031)}&\emph{0.001 (0.026)}&\emph{0.001 (0.023)}&\emph{0.001 (0.000)}\\
PERCBLK&\emph{-0.001 (0.001)}&\emph{-0.001 (0.000)}&\emph{-0.0005 (0.000)}&\emph{0.001 (0.000)}&\emph{0.002 (0.000)}&\emph{0.001 (0.000)}\\
BLKDIFF&0.000 (0.134)&0.000 (0.380)&&\emph{-0.002 (0.001)}&\emph{-0.002 (0.000)}&\emph{-0.002 (0.000)}\\
MEDINC&0.000 (0.199)&0.000 (0.126)&0.0000 (0.147)&\emph{0.000 (0.031)}&\emph{0.000 (0.030)}&\emph{0.000 (0.019)}\\
MEDINCRAT&\emph{0.019 (0.034)}&\emph{0.021 (0.001)}&\emph{0.0174 (0.000)}&\emph{0.077 (0.001)}&\emph{0.078 (0.001)}&\emph{0.081 (0.001)}\\
WHTDIFF&0.000 (0.395)&&&0.000 (0.901)&&\\
POPDENS&0.003 (0.396)&&&-0.008 (0.350)&-0.009 (0.301)&\\
POPDENSRAT&-0.006 (0.195)&-0.005 (0.239)&&0.015 (0.285)&0.016 (0.255)&\\
POPRATIO&-0.011 (0.799)&&&-0.016 (0.905)&&\\
Adj. R-sq.&0.3223&0.3265&0.3229&0.3255&0.3319&0.3222\\
N&310&310&310&307&307&307\\

\hline
\end{tabular}
\end{center}
\caption*{\emph{Significant at 5 percent.}}
\end{adjustwidth}
\end{table}

Controlling for overall diversity levels (both the suburbs’ H index values and their overall differences with the core city), as well as the length and percentage of the common border, there are two key significant determinants. The ratio of median income is significantly positive, indicating that richer suburbs are more likely to have high BDI maxima (with less diversity than their core cities). While suburbs’ median income levels, and the differences versus the core city in the percentages of White and Black residents are not significant, suburbs’ percentages of Black residents are significantly negatively associated with the BDI maximum. Border effects are weaker if suburbs have more African-American residents, suggesting that discriminatory real-estate practices might be partially responsible for the disparities discussed in this study. None of the population variables are significant. 

Repeating these regressions using the maximum values of the BDI that measures disparities in the percentage Black, income (both absolute and relative to the core city) is positively related to this index. Richer suburbs, therefore, are more likely to have larger disparities and, most likely, to have lower percentages of Black residents. This model explains about one-third of the variance; further research would be necessary, however, to incorporate additional explanatory variables and to uncover the specific processes behind these findings.

\section{Conclusion}
Many large U.S. cities have sharp discontinuities in the ethnic makeup of neighborhoods on either side of the municipal border. These often can be attributed to differences in zoning or school quality, or may be due to a legacy of housing or other forms of discrimination. But while these disparities are often known to locals, and sometimes become nationally known, no empirical study has thus far attempted to isolate them quantitatively. This study does so, developing a so-called “Border Disparity Index” for ethnic diversity in U.S. census block groups along the borders of 25 major U.S. cities and their suburbs. These disparities are then examined at the metropolitan, as well as the place, level.

	Overall, this study shows that the index created here captures patterns that are both visually apparent and, in the test case of Milwaukee, are identified using traditional cluster analysis methods. These results confirm that this new measure provides a valid method of assessing the effects we wish to examine here. This study arrives at interesting conclusions for our set of metros, core cities, and individual suburbs. In particular, metros such as Atlanta, Detroit, and Philadelphia have large disparity indexes, particularly within the core cities themselves. Diversity is therefore relatively high on the suburban side of the city border in these areas. Chicago shows itself to be a highly diverse metropolitan area; its index value is exceptionally high, and its highest overall value also falls within the city itself. In some places, suburban diversity is high; and in others, the city side of the border is more diverse. The opposite is true for cities such as Portland, with low index values in all categories, suggesting a more homogenous metropolitan area. The fact that much (but not all) of these significant findings can be found in the Northeast and Midwest is worthy of further investigation.

An examination of individual suburbs shows the highest disparity index values to be located within suburbs of Chicago and Milwaukee, with the Indianapolis metro home to a number of low-diversity suburbs. High-diversity suburbs, on the other hand, can be found near Detroit, Philadelphia, and Buffalo, among other cities, possibly reflecting monoethnic populations (often Black), who live in parts of the city that touch the border, but not in the suburbs themselves. Chicago also has a number of high-diversity suburbs as well, and Los Angeles is particularly interesting in that it keeps diversity in versus some suburbs, and out versus others.

A regression model for 310 of these suburbs shows that, controlling for overall diversity levels and border lengths, the suburban/core city income ratio is significantly correlated with place-wide BDI maximum values. Suburbs’ percentages of black residents are negatively correlated. This indicates that, all else equal, higher-income suburbs are more likely to keep diversity “out,” while the opposite is true for suburbs with more Black residents. 

While more needs to be done to investigate the disparities uncovered here, including the role of physical barriers such as those proposed by Mitchell and Lee (2013),  these findings can be useful for policymakers and community members. Housing policy might be addressed to mitigate disparities in specific areas where disparities are highest. Resources can be directed at the city and suburban levels as well. In this way, segregation can be addressed once its effects are isolated geographically.   
 
\nocite{Anacker2017}
\nocite{Chakravorty1996}
\nocite{Downey2003}
\nocite{Hero2016}
\nocite{Jargowsky1996}
\nocite{Lichter2015}
\nocite{Massey1998}
\nocite{South1997}
\nocite{South2011}
\nocite{Dean2018}
\nocite{Mitchell2013}
\nocite{Siegel2013}
\nocite{Logan2011}
\nocite{Harris2014}

\bibliographystyle{abbrv}
\bibliography{BorderPaper}

\end{document}